# A systematic search for step-like anomalies in the tritium $\beta$-decay spectrum in the Troitsk-$\nu$-mass experiment

A.V.Lokhov[a], F.V.Tkachov[b], P.S.Trukhanov[a]

[a] Department of Physics, Moscow State University, Moscow, 119991
[b] Institute for Nuclear Research RAS, Moscow, 117312

*Abstract.*
The issue of step-like anomalies in the tritium $\beta$-decay spectrum as measured in the Troitsk-$\nu$-mass experiment is addressed in the context of the new analysis in a systematic fashion using efficient statistical tests specifically derived for the purpose.
It is concluded that the presence of the anomaly cannot be statistically asserted with a high confidence level.

## 1. Introduction

The recent report [1] presented a reanalysis of the data of the Troitsk-$\nu$-mass experiment [2] (some details of the experimental setup and a brief description of the measured cumulative spectrum can be found in [1]). It has yielded an upper limit on the electron neutrino mass that slightly improves upon both the previous analysis [3] and the results of the Mainz expeiment [4].

Although the new analysis per se does not seem to have any irregularities, one cannot ignore the fact that the previous analysis [3] found a (step-like) anomaly in the data that was a subject of discussion for a number of years.

It therefore seems appropriate to devote a dedicated investigation of the issue of the step-like anomaly as a sort of postscriptum to the analysis of [1], having in view that various physical explanations of the anomaly were proposed and it would be nice to know that the anomaly has not simply went below the radar of the non-specific $\chi^2$ criterion, but is reliably excluded by a dedicated statistical test.

Direct measurements of neutrino mass in the tritium $\beta$-decay (for a review see [5]) are model-independent unlike neutrino mass estimates from cosmology [6] and neutrinoless double $\beta$-decay searches [7]. On the other hand, the neutrino oscillation experiments provide the differences between the squared neutrino masses (and the mixing angles) [8]-[16] but not the absolute masses.

Neutrino mass measurements in the tritium $\beta$-decay involve a precise investigation of the endpoint region of the energy spectrum of the decay electrons, which is the most sensitive part of the spectrum with respect to the neutrino mass.

Therefore, any deviations (either statistical or due to some physical effects) near the endpoint region can contribute significantly to the estimate of the neutrino mass squared. This issue arose in the first analysis [3] of the Troitsk-$\nu$-mass data that yielded an estimate of the neutrino mass squared well below zero: $m_\nu^2 \sim -(10 \div 20)\ eV^2$. This was explained by an apparent excess of electrons near the end-point energy, and various physical explanations for this were suggested (additional week interactions [17] and tachyonic neutrinos [18]). In the Troitsk-$\nu$-mass cumulative spectrum such an excess takes the form of a step described by two parameters: height and position. Then an acceptable value $m_\nu^2 = -2.3 \pm 2.5_{fit} \pm 2.0_{syst}\ eV^2$ was obtained.

The new analysis [1] exploits the method of quasi-optimal weights [19] that took into account the non-Gaussian nature of the data, and an improved theoretical model of the experimental setup. Most importantly, the source sickness parameter values were reexamined in [1] for each run, which proved to be the main reason for the apparent disappearance of the anomaly in the new analysis, with the resulting estimate of the neutrino mass squared $m_\nu^2 = -0.67 \pm 1.89_{fit} \pm 1.68_{syst}\ eV^2$.

However, as we have pointed out at the beginning of this text, the fact of an anomaly seen in the previous analysis [3] warrants an extra effort because neither the least squares used in the first analysis nor the method of quasi-optimal weights in the new analysis are sensitive enough to the anomalies in the spectrum.

Ref. [20] demonstrated how the general flexible approach behind the method of quasi-optimal weights allows one to derive statistical criteria tuned to anomaly searches, specifically to the search of step-like anomalies, that by construction are — unlike the non-specific $\chi^2$ test — powerful and efficient in the usual statistical sense.

In particular, two convenient dedicated criteria for the step-like anomaly in the tritium $\beta$-decay cumulative spectrum were constructed (the «quasi-optimal» test and the criterion of «pairwise neighbours' correlations»). Both were shown to be more statistically efficient for the purpose than the conventional ones (the $\chi^2$ and a Kolmogorov-Smirnov type criteria).

However, a further technical issue — the one addressed in the present paper — is that the Troitsk-$\nu$-mass data as used in [1] consist of eleven runs. Each run can be treated with the two tests mentioned above, and then for each test one arrives at a set of 11 numbers that have to be statistically integrated in order to arrive at a single informative number. This is done by observing that in the absence of the anomaly (the null-hypothesis), the eleven numbers for each test form a random sample with a known probability distribution, and this fact can be made use of.

Before we turn to the actual calculations, we note that our present analysis can be viewed from a slightly different angle. Indeed, if one ignores the issue of anomalies, our analysis becomes an additional test of the correctness of estimates of the source thickness, a parameter that proved to play a key role in the new analysis.

## 2. The experimental data

The data set of each run contains a set of values of energies (or retarding potentials controlled by the experimentalists) $E_i$, $i = 1,..,M$ (here and below, $M$ is the number of different energy points at which the measurements are carried out). Each energy corresponds to the experimental number of events $N_i = N(E_i)$ and the fit $\mu_i = \mu(E_i)$.

The $N_i$ are Poisson-distributed with the mean values estimated by $\mu_i$. So one has eleven sets of triplets $\{E_i, N_i, \mu_i\}_q$, $q = 1,..,11$ for the eleven runs of the Troitsk-$\nu$-mass (the number of experimental points $M$ is usually 44 but it can slightly change from run to run due to a pre-processing of the data).

The special criteria constructed in ref. [20] are statistically efficient for searches of step-like anomalous contributions: $\mu'(E) = \mu(E) + \Delta \cdot \theta(E_{st} - E)$, where $E_{st}$ stands for the position of the step, $\Delta$ is the height of the step (a step in the spectrum emerges due to integration of a $\delta$-shaped anomaly). Note that in the new data analysis the values of fit $\mu(E_i)$ are obtained without assuming a step in the spectrum (i.e. $\Delta = 0$).

Generally speaking, the position of the step is unknown. In addition, the early data analysis indicated that the position varies with time in the approximate interval $18558\ eV \leq E_{st} \leq 18568\ eV$ [3]. One of the constructed tests, the quasi-optimal one, depends on a parameter $E_0$ which we choose to be the centre point of this interval. This ensures a maximal sensitivity of this test to the anomalies with $E_{st}$ in the above interval. The criterion of the «pairwise neighbours correlations» is parameter-free but rather more powerful in our case than the $\chi^2$ test.

## 3. Evaluation of the criteria

First of all one obtains the experimental values of the two criteria for each of the eleven runs of the Troitsk-$\nu$-mass similarly to the example given in [20]. Then the set of eleven numbers for each test is examined statistically.

The statistics of the «quasi-optimal» criterion according to [20] is

$$S_{q-opt} = \sum_{i=1}^{M} w_i \cdot \xi_i, \qquad (1)$$

where $\xi_i = (N_i - \mu_i)/\sqrt{\mu_i}$, and $w_i$ are chosen as:

$$w_i = \begin{cases} (m-i)/m, & i \leq m, \\ (m-i)/(M-m), & i > m, \end{cases} \qquad (2)$$

where $m$ is defined via the inequality $E_m \leq E_{st} < E_{m+1}$.

The statistics of the criterion of «pairwise neighbours correlations» is

$$S_{pair} = \sum_i \xi_i \cdot \xi_{i+1} \qquad (3)$$

The experimental values $N_i$ are used to calculate the corresponding experimental values of the criteria $S_{q-opt}^q$ and $S_{pair}^q$ for each run $q$.

The distribution functions $F_1(S_{q-opt})$ and $F_2(S_{pair})$ for the two criteria are constructed via Monte-Carlo modelling (by generating the spectrum $\{N_i\}$, assuming that the height of the step $\Delta$ is 0, evaluating the values $S_{q-opt}$ and $S_{pair}$ for each set $\{N_i\}$, and then using the sets $\{S_{q-opt}\}$ and $\{S_{pair}\}$ thus obtained to construct the distribution functions).

Then one determines the values of these functions $F_1(S_{q-opt})$ and $F_2(S_{pair})$ at the experimental points:

$$\alpha^q_{q-opt} = F_1(S^q_{q-opt}), \quad \alpha^q_{pair} = F_2(S^q_{pair}) \qquad (4)$$

Each set $\{E_i, N_i, \mu_i\}_q$ yields the two values $\alpha^q_{q-opt}$ and $\alpha^q_{pair}$, thus we have 22 results presented in Table 1.

| Run # | 22 | 23 | 24-1 | 24-2 | 25 | 28 | 29 | 30 | 31 | 33 | 36 |
|---|---|---|---|---|---|---|---|---|---|---|---|
| $\alpha^q_{q-opt}$ | 0.382 | 0.359 | 0.381 | 0.522 | 0.478 | 0.266 | 0.510 | 0.371 | 0.365 | 0.570 | 0.207 |
| $\alpha^q_{pair}$ | 0.883 | 0.571 | 0.471 | 0.604 | 0.920 | 0.829 | 0.141 | 0.994 | 0.113 | 0.702 | 0.810 |

**Table 1.** The values of the distribution functions of the two tests for the eleven runs of the Troitsk-$\nu$-mass experiment.

If one chooses the 95% C.L. (the confidence level is the probability to accept the null-hypothesis while it is correct; the corresponding confidence probability is $\alpha_0 = 0.95$) then the null-hypothesis ($\Delta = 0$) is rejected if $\alpha^q > \alpha_0 = 0.95$. The table shows that this occurs only in one case: the criterion of «pairwise neighbours correlations» rejects the null-hypothesis in run 30 at the 95% C.L. as it yields the value $\alpha^q_{pair} = 0.994 > a_0 = 0.95$.

## 4. Uniformity analysis

All the values $\alpha$ in each row of Table 1 are independent. On the other hand, the variates $F(S^q)$ are distributed uniformly on the interval [0,1]. Therefore, one can perform goodness-of-fit tests for the two samples $\{\alpha^q_{q-opt}\}_q$ and $\{\alpha^q_{pair}\}_q$ with respect to the uniformity hypothesis. Below we list our results for seven such tests and two symmetry tests, for which we found reliable formulae with tables of critical levels.

Note that the two samples are small and one can not expect highly reliable results here. We will use the 95% C.L.

### 1) $\chi^2$ test [21]

Here one divides the segment [0, 1] into $k$ equal intervals and calculates the quantity

$$\chi^2 = \sum_1^k \frac{(\nu_i - N \cdot p_i)^2}{N \cdot p_i}, \qquad (5)$$

where $p_i = k^{-1}$ is the probability of $\alpha$ to hit the interval number $i$, $\nu_i$ is the number of points $\alpha$ that actually fall into the $i$-th interval. The number of intervals $k$ should satisfy $N/k \geq 5$ [21]. In our case $N = 11$ and $k = 2$. In this case the calculated variate (5) has the $\chi^2$ distribution with one degree of freedom.

For both $\{\alpha_{q-opt}\}$ and $\{\alpha_{pair}\}$ one obtains $\chi^2 = 2.273$. The critical value for the 95% C.L. here is 3.841 [22]. Thereby the hypothesis of uniform distribution of $\alpha_{pair}$ and $\alpha_{q-opt}$ is not rejected at 95% confidence level.

## 2) Kolmogorov-Smirnov test [21]

This test is based on the comparison of the cumulative distribution functions - the empirical and theoretical one (see Fig.1). (Recall that we are dealing with the uniform distribution.)

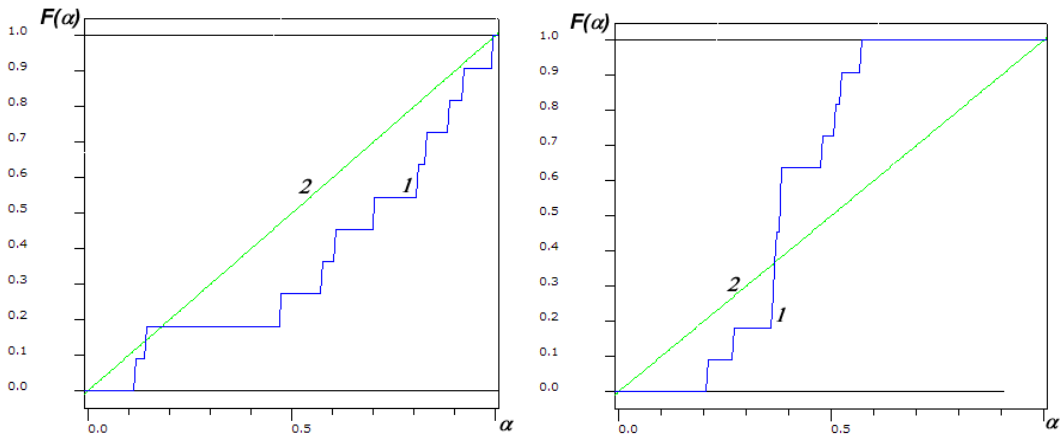

**Figure 1.** The empirical (1) and theoretical (the straight line, 2) cummulative distribution functions for $\alpha_{pair}$ (left) and $\alpha_{q-opt}$ (right).

The statistics of the criterion is

$$D_n = \operatorname{Sup}_x |S_n(x) - F(x)|, \qquad (6)$$

where $n$ is the size of the sample $\{\alpha\}$, $S_n(x)$ is the empirical distribution function of the variate $\alpha$ and $F(x)$ is the corresponding theoretical distribution function.

The calculation of the value $D_n$ for the sample $\{\alpha_{pair}\}$ yields $D_{11} = 0.299$, and for the sample $\{\alpha_{q-opt}\}$ — $D_{11} = 0.430$. The critical value at the 95% C.L. is 0.391. So the hypothesis that $\alpha_{q-opt}$ is distributed uniformly is rejected, while the uniformity of $\alpha_{pair}$ is accepted.

The following six criteria are specific to the uniform distribution on [0, 1].

## 3) Sherman criterion [23]

The statistics of the Sherman test has the following form:

$$\omega_n = \frac{1}{2}\sum_{i=1}^{n+1}\left|\alpha_i - \alpha_{i-1} - \frac{1}{n+1}\right|, \qquad \alpha_0 = 0; \quad \alpha_{n+1} = 1; \qquad (7)$$

($\alpha_i$ are sorted in the ascending order, i.e. $\alpha_i$ is the order statistics). The value of the criterion is $\omega_{11} = 0.334$ for the sample $\{\alpha_{pair}\}$ and $\omega_{11} = 0.492$ for $\{\alpha_{q-opt}\}$. The critical value (with confidence probability 0.95) is 0.469. Therefore the uniformity of $\{\alpha_{pair}\}$ is accepted, but for $\{\alpha_{q-opt}\}$ it is rejected.

**4) Moran criterion [23]**

The statistics of the Moran test is

$$M_n = -\sum_{i=1}^{n+1}\ln\left[(n+1)D_i\right], \qquad D_i = \alpha_i - \alpha_{i-1}, D_1 = \alpha_1, D_{n+1} = 1 - \alpha_n. \qquad (8)$$

The hypothesis is accepted if $M_1(0.95) \leq M_n \leq M_2(0.95)$; with $n = 11$ the corresponding 95% critical values are $M_1(0.95) = 2.704$ and $M_2(0.95) = 11.375$. The calculated values are $M_{11} = 13.013$ for $\{\alpha_{q-opt}\}$ and $M_{11} = 5.243$ for $\{\alpha_{pair}\}$. The hypothesis of uniformity is thus accepted for $\alpha_{pair}$ and rejected for $\alpha_{q-opt}$.

**5) Cheng-Spiring test [23]**

The statistics of the test is

$$W_n = \left[(\alpha_n - \alpha_1)\frac{n+1}{n-1}\right]^2 \Big/ \sum_{i=1}^{n}(\alpha_i - \bar{\alpha})^2. \qquad (9)$$

The hypothesis is accepted if $W_1(0.95) \leq W_n \leq W_2(0.95)$. The calculated value of the criterion $W_{11} = 1.25$ for the sample $\{\alpha_{pair}\}$ lies between the 95% C.L. critical values: $0.92 \leq 1.25 \leq 1.73$; thus the uniformity of $\alpha_{pair}$ is not rejected. The value of the criterion for the sample $\{\alpha_{q-opt}\}$ is $W_{11} = 1.56$, $0.92 \leq 1.56 \leq 1.73$, so the uniformity of $\alpha_{q-opt}$ is also accepted.

**6) Frocini test [23]**

The statistics of the Frocini criterion for the uniform distribution is

$$B_n = \frac{1}{\sqrt{n}}\sum_{i=1}^{n}\left|\alpha_i - \frac{i-0.5}{n}\right|. \qquad (10)$$

A distribution is acknowledged as uniform if $B_n < B_n(0.95)$.

The calculated value for the sample $\{\alpha_{pair}\}$ is $B_{11} = 0.464$, the corresponding critical value is $B_{11}(0.95) = 0.5806$. So the distribution of $\alpha_{pair}$ is acknowledged as uniform. For the second sample the criterion yields $B_{11} = 0.611 > 0.5806$, rejecting the uniformity of the $\alpha_{q-opt}$ distribution.

## 7) Yang criterion [23]

The Yang test statistics is

$$M = \sum_{i=1}^{n} \min(D_i, D_{i+1}), \qquad D_1 = \alpha_i, D_i = \alpha_i - \alpha_{i-1}, D_{n+1} = 1 - \alpha_n \qquad (11)$$

The uniformity is accepted if $M_1(0.95) \leq M_n \leq M_2(0.95)$. For the sample $\{\alpha_{pair}\}$ one obtains $0.3 \leq 0.435 \leq 0.6$ and the uniformity is not rejected at the 95% C.L. Applying the Yang test to the sample $\{\alpha_{q-opt}\}$, one obtains $M_{11}^{q-opt} = 0.241 < M_1(0.95)$. So the Yang test rejects the hypothesis of the uniform distribution of the variate $\alpha_{q-opt}$.

## Symmetry tests

The uniform distribution on the segment [0, 1] is symmetrical relative to the point $a = 0.5$. Therefore one can apply symmetry tests to the samples $\{\alpha_{pair}\}$ and $\{\alpha_{q-opt}\}$. It is important to use tests that are applicable to small samples.

### a) Sign criterion [23]

The statistics of the sign test is the quantity $K = \min(K^+, K^-)$, where $K^+$ and $K^-$ are the numbers of positive and negative values of $y_i = \alpha_i - 0.5$. The symmetry hypothesis is rejected if $K = \min(K^+, K^-) < K(\alpha)$. For both samples $\{\alpha_{pair}\}$ and $\{\alpha_{q-opt}\}$ the test yields $K = 3$, while the critical value for the confidence probability 0.95 is 2. Therefore, the symmetry hypothesis is accepted for both samples.

### b) Single-sample Vilkokson test [23]

The statistics of this test for the sample of the size of $n$ elements is

$$T^+ = \sum_{i=1}^{n} R_i^+, \qquad (12)$$

where $R_i^+$ is the number of the quantity $|y_i| = |\alpha_i - 0.5|$ in the ordered sequence $|y_1| \leq |y_2| \leq ... \leq |y_n|$, provided $y_i > 0$. The calculated value for the sample $\{\alpha_{pair}\}$ is $T^+ = 1+3+4+5+6+8+9+10+11 = 49 < T^+(0.95) = 51$, and for the sample $\{\alpha_{q-opt}\}$ the test yields $T^+ = 58 > T^+(0.95) = 51$. Thus the hypothesis of symmetry of the distribution of the variate $\alpha_{q-opt}$ is rejected at the 95% C.L. At the same time this test does not reject the symmetry of the $\alpha_{pair}$ distribution.

## 5. Summary and conclusions

We have applied the special criteria to the search of step-like anomalies in the cumulative tritium $\beta$-decay spectrum. We used the data from the eleven runs of the Troitsk-$\nu$-mass experiment.

The tests of the hypothesis that the anomaly is present in the spectrum, with the «quasi-optimal» criterion and the criterion of «pairwise neighbours correlations» show that only in one run (run #30) the value of one of the two criteria (namely, the «pairwise neighbours correlations» test) exceeds the critical value for the 95% C.L. Only in this single case the no-step hypothesis is rejected. Note that the no-step hypothesis is rejected in this case at the 99% C.L as well (note that the other criterion sees nothing here). The overall effect of this single point can only be meaningful in the statistical context of the entire set of eleven numbers (see Table 2); we verified that the omission of this point has no appreciable effect on the results in Table 2.

To search for rough anomalies in the spectrum we have also studied the two sets of variates (the two samples of values of the distribution function of each criterion at the corresponding experimental points). In theory these variates are distributed uniformly and one can test the distribution of each sample for uniformity and symmetry. The results of these uniformity and symmetry goodness-of-fit tests are summarized in Table 2, where we also added the 99% confidence level boundaries in addition to the 95% ones cited in the text above.

| Criterion | $\alpha_{q-opt}$ | $\alpha_{pair}$ | Critical region 95% | Critical region 99% |
|---|---|---|---|---|
| $\chi^2$ | 2.273 | 2.273 | <3.841 | <6.635 |
| Kolmogorov | 0.430 | 0.299 | <0.391 | <0.468 |
| Sherman | 0.492 | 0.334 | <0.469 | <0.521 |
| Moran | 13.013 | 5.243 | [2.704, 11.375] | [1.81, 14.21] |
| Cheng-Spiring | 1.56 | 1.25 | [0.92, 1.73] | [0.83, 1.99] |
| Frocini | 0.611 | 0.464 | <0.581 | <0.749 |
| Yang | 0.241 | 0.435 | [0.3, 0.6] | [0.2, 0.6] |
| Sing | 3 | 3 | >1 | >0 |
| Vilkokson | 58 | 49 | <51 | <58 |

**Table 2.** The summary of uniformity and symmetry criteria for the samples of the eleven experimental values of the two special criteria.

At the 95% C.L. the hypothesis of uniformity of the sample $\{\alpha_{q-opt}\}$ (which corresponds to the «quasi-optimal» criterion) is rejected by the Yang test, Kolmogorov-Smirnov criterion, Moran, Sherman and Frocini tests. The uniformity of $\{\alpha_{q-opt}\}$ is accepted at the 95% C.L. by the $\chi^2$ and the Cheng-Spiring tests. The uniformity of the sample $\{\alpha_{pair}\}$ (the criterion of «pairwise neighbours correlations») is confirmed by all tests.

The symmetry of the $\{\alpha_{q-opt}\}$ sample is rejected only by the Vilkokson test, but is accepted by the simple sign test. At the same time the symmetry of $\{\alpha_{pair}\}$ distribution is accepted by both criteria at the 95% C.L.

At first glance it might seem that one set of numbers (the eleven numbers for the eleven runs) was simply replaced by another (the nine numbers for the nine uniformity tests; and on top of everything, these latter numbers do not form a random sample from a known statistical ensemble). However, each of the latter numbers is informative by design, i.e. is meant to explicitly convey information on the correspondence of the initial random sample of the eleven numbers to the null-hypothesis (the absence of the step-like anomaly).

Unfortunately, with so many uniformity and symmetry criteria available, it is statistically likely that some of them would yield numbers outside the critical region (although it is impossible to say anything more specific because, again, the numbers do not form a random sample amenable to statistical analysis). This indeed appears to be the case with the 95% critical region. One could argue that in such a situation a higher confidence level should be chosen, e.g. 99%. Then, indeed, the rightmost column of the table shows that all the uniformity and symmetry criteria fall within the 99% critical region, and one concludes that there is no statistical ground to claim that an anomaly is present.

## Acknowledgements

The authors thank the members of the Troitsk-$\nu$-mass experiment for kindly providing the bits of data relevant for the present study prior to completion of the analysis [1] (A.A.Nozik's help in this regard is explicitly acknowledged). Special thanks are due to A.I. Berlev for his invariable interest in this work, V.S.Pantuev for reading parts of the text and N.A.Titov for stimulating criticisms. We also thank members of the Department of theoretical physics of INR RAS for discussions.